\documentclass[intlimits,twoside,a4paper]{article}
\usepackage[cp1251]{inputenc}

\usepackage{multirow}

\usepackage[eqsecnum]{cmpj3}

\usepackage{bm}

\issue{2021}{24}{1}{13702}
\doinumber{10.5488/CMP.24.13702}
\title[Structural,  elastic, electronic and optical properties]%
{First principle study of structural,  elastic, electronic and optical properties of Pb$_{0.5}$Sn$_{0.5}$TiO$_{3}$ and Pb$_{0.5}$Sn$_{0.5}$Ti$_{0.5}$(Zr$_{0.5}$)O$_{3}$} 

\author[S.G. Kuma, M.M. Woldemariam]{S.G. Kuma\refaddr{label1},
         M.M. Woldemariam\refaddr{label2}}
\addresses{
\addr{label1} Department of Physics, Wollega University, P.O.Box 395, Nekemte, Ethiopia
\addr{label2} Department of Physics, Jimma University, P.O.Box 378, Jimma, Ethiopia
}

\date{Received May 29, 2020, in final form August 28, 2020}
\begin{document}

\maketitle

\begin{abstract}
The structural, electronic, elastic and optical properties of tetragonal (P4mm) phase of Pb$_{0.5}$Sn$_{0.5}$TiO$_{3}$ (PSTO) and Pb$_{0.5}$Sn$_{0.5}$Ti$_{0.5}$(Zr$_{0.5}$)O$_{3}$ (PSTZO) are examined by first-principles calculations based on the density functional theory (DFT) using the pseudo-potential plane wave (PP-PW) scheme in the frame of generalized gradient approximation (GGA). We have calculated the ground state properties such as equlibrium lattice constants, volume, bulk modulus and its pressure derivative. From elastic constants, mechanical parameters such as anisotropy factor, elastic modulus and Poisson's ratio are obtained from the Voigt-Reuss-Hill average approximation. Rather than their averages, the directional dependence of elastic modulus, and Poisson's ratio are modelled and visualized in the light of the elastic properties of both systems. In addition, some novel results, such as Debye temperatures, and sound velocities are obtained. Moreover, we have presented the results of the electronic band structure, densities of states and charge densities. These results were in favourable agreement with the existing theoretical data. The optical dielectric function and  energy loss spectrum of both systems are also computed.  Born effective charge (BEC) of each atoms for both systems is computed from functional perturbation theory (DFPT). Finally, the spontaneous polarization is also determined from modern theory of polarization to be 0.8662~C/m$^{2}$ (PSTO) and 1.0824~C/m$^{2}$ (PSTZO). 
\end{abstract}

\section{Introduction}
Perovskite (ABO$_{3}$) has held the interest of crystallography for a significant period of time due to the wide range of compositions and the existence of unique properties such as ferroelectricity. The development of perovskite type materials exhibiting a very high dielectric constant and spontaneous polarization is of considerable interest since they play an important role in electronics /microelectronics and have various technological applications \cite{rodel2009perspective,kittel1969einfuhrung,dove1997theory,resta1994macroscopic,knapp2006site,jaffe2012piezoelectric}. Most of the ferroelectric materials used for device applications are Pb-based such as lead titanate (PbTiO$_{3}$), lead zirconate titanate Pb(Zr$_{x}$Ti$_{1-x}$)O$_{3}$ (PZT) ($x$ = composition), lead lanthanum zirconate titanate (Pb$_{1-x}$La$_{x}$)(Zr$_{y}$ Ti$_{1-y}$)$_{1-x/4}$O$_{3}$ (PLZT) and lead magnesium niobate Pb(Mg, Nb)O$_{3}$ (PMN) \cite{sani2004high,noheda2000stability,noheda2002phase,wang2014first}. Nowadays, there is observed  environmental as well as health concern regarding the toxicity of lead based oxides which are volatile during processing. The volatilization of toxic lead oxide during high-temperature sintering not only causes environmental pollution but also generates instability in composition and affects the electrical properties of the products. Moreover, the products containing Pb-based gadgets are not recyclable \cite{bell2018lead}.  An increased environmental awareness over the past 20 years has resulted in several governmental regulations world-wide that are putting increasingly stringent requirements on the use of materials in electronic devices \cite{robinson2009waste,babu2007electrical}. Consequently, this has opened the searches on identifying new and more environmentally friendly ferroelectric materials and other alternative compounds as well as on seeking novel Pb-free ferroelectric materials with cation displacements of comparable (or greater) magnitude for their application in the future sustainable electroactive materials \cite{taib2014first,zhao2017lead,coondoo2013lead}. 
The total replacement of Pb-based materials in technological devices remains almost improbable because of the unsatisfactory performance of other materials. However, modification efforts to reduce the consumption of toxic Pb$^{2+}$, such as by substitution or doping techniques, remain necessary. As reported by Taib et al. \cite{taib2014first} in order to decrease the harm of Pb$^{2+}$ element, the potential elements such as Sn$^{2+}$ and Ge$^{2+}$ are eventually expected to be more environmental friendly materials. Cohen and Ganesh \cite{cohen2011class} reported that a novel compound of Pb$_{0.5}$Sn$_{0.5}$ZrTiO$_{3}$ (PSZT) has much higher values of piezoelectric coefficients than PZT. Roy and Vanderbilt \cite{roy2011theory} also explained that strong ferroelectric properties could exist in double rock-salts PbSnZrTiO$_{6}$. By substitution reducing the concentration of Pb$^{2+}$ we mainly focus on the investigation of structural, elastic, electronic, spontaneous polarization and optical properties of PSTO  and  PSTZO compounds. 
The paper is organized as follows. In section 2, the techniques for the calculations are described, and the computational details are presented. In section~3, the present results of the structural, electronic, elastic, and optical properties for both systems are discussed and compared with the available previous results. Finally, the conclusion of the present work is drawn in the last section.

\section{Computational details}

\begin{figure}[!b]
\centering
a)
    {\includegraphics[height=4cm, width=3cm]{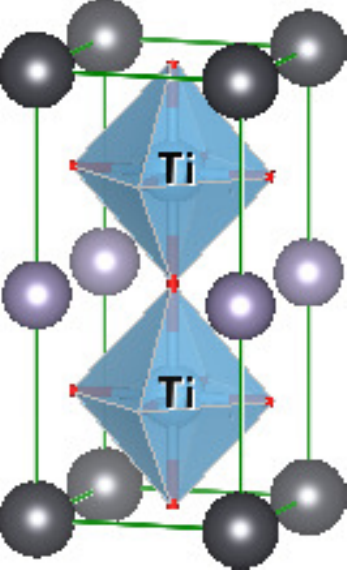}}
      \hspace{1cm}
  b)
    {\includegraphics[height=4cm, width=3cm]{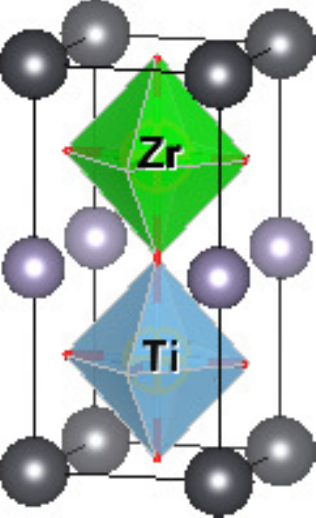}}
  \caption{(Colour online) The crystal structure of a) PSTO and b) PSTZO.}
  \label{figure1}
\end{figure}

The first principles calculations were conducted using the density functional theory (DFT) as implemented in Quantum Espresso software Package (QE) \cite{giannozzi2009quantum} open source code within General Gradient Approximation (GGA) functional \cite{perdew1996generalized}. The atomic crystal structures of supercell $1{\times}1{\times}2$ of PSTO and PSTZO were designed in tetragonal (P4mm, 99 space group) phase for the composition of Ti/Zr 50/50 and Pb/Sn  50/50 as illustrated in figure~\ref{figure1}. The direct substitutive point defect techiques were used to replace Pb by Sn and Ti by Zr atom.  The ultra-soft pseudopotentials \cite{hasnip2006electronic} were used to treat the interaction of the electrons with the ion cores, where Pb (6s, 6p), Sn (5s, 5p), Ti (4p, 3d), Zr (4d, 5s) and O (2s, 2p) electrons were treated as valence states. A plane-wave cut-off energy of 80~Ry and the Brillouin zone with $4\times4\times4$ Monkhorst Pack \cite{monkhorst1976special} k-point were applied. Geometric optimization was performed under the convergence criteria: energy $1.0\cdot 10^{-4}$, force $1.0\cdot 10^{-3}$~Ry/Bohr, cell $5.0\cdot 10^{-1}$~kbar. Density functional perturbation theory (DFPT) \cite{gonze1997dynamical} was used to calculate the Born effective charge (BEC) of the systems. In addition, the spontaneous polarization was computed using the Berry-phase approach \cite{resta2007theory}. Using this approach, the total polarization $P$ for a given crystalline geometry can be calculated as the sum of ionic and electronic contributions.

\section{Result and discussions}
\subsubsection{Structure properties}
\begin{figure}[!t]
\centering
{\includegraphics[width=0.64\textwidth]{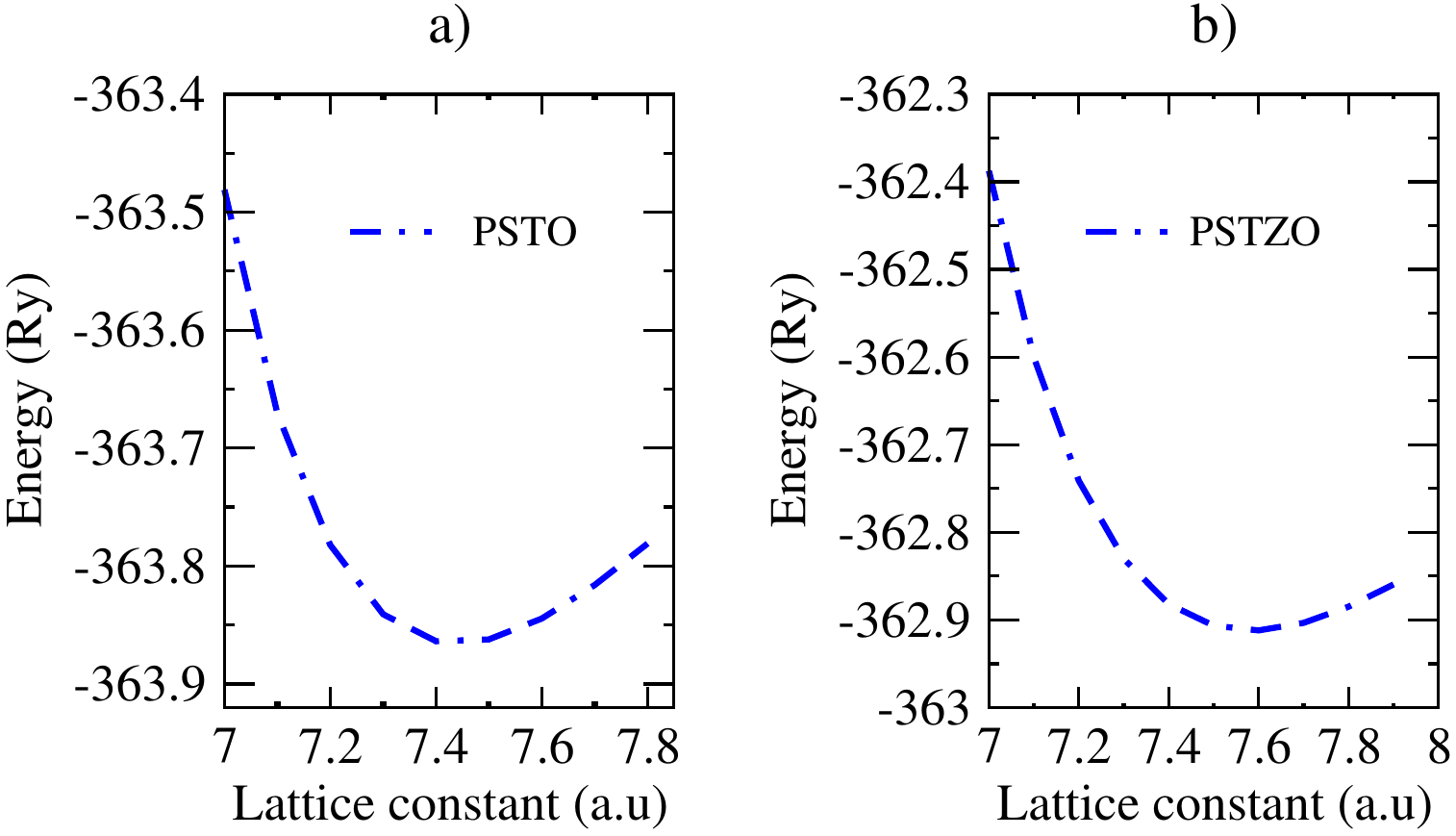}}
\caption{(Colour online) The energy with respect to lattice constant of a) PSTO and b)
PSTZO.} \label{figure2}
\end{figure}

The equilibrium lattice constant is a fundamental part of the structural information related to crystal structures. To calculate the theoretical lattice constant, the cutoff energy and k-point sampling obtained from total energy convergence test are employed. In our computations, the lattice constant was varied with 0.1~Bohr radius increments. From figure~\ref{figure2}, the optimized equilibrium lattice  constant of PSTO is $a = 3.978$~\AA ~$\text{and}$ $c= 8.353$~\AA~and also the optimized equilibrium lattice constant of PSTZO is $a$ = 4.025~\AA~and $c$ = 8.610~\AA,~which are in better agreement with the previous theoretical facts \cite{resta2007theory, hussin2017theoretical, wang2010synthesis}. The computed lattice constants and its atomic positions are significant factors of the material stability.

\begin{figure}[!t]
\centering
{\includegraphics[width=0.64\textwidth]{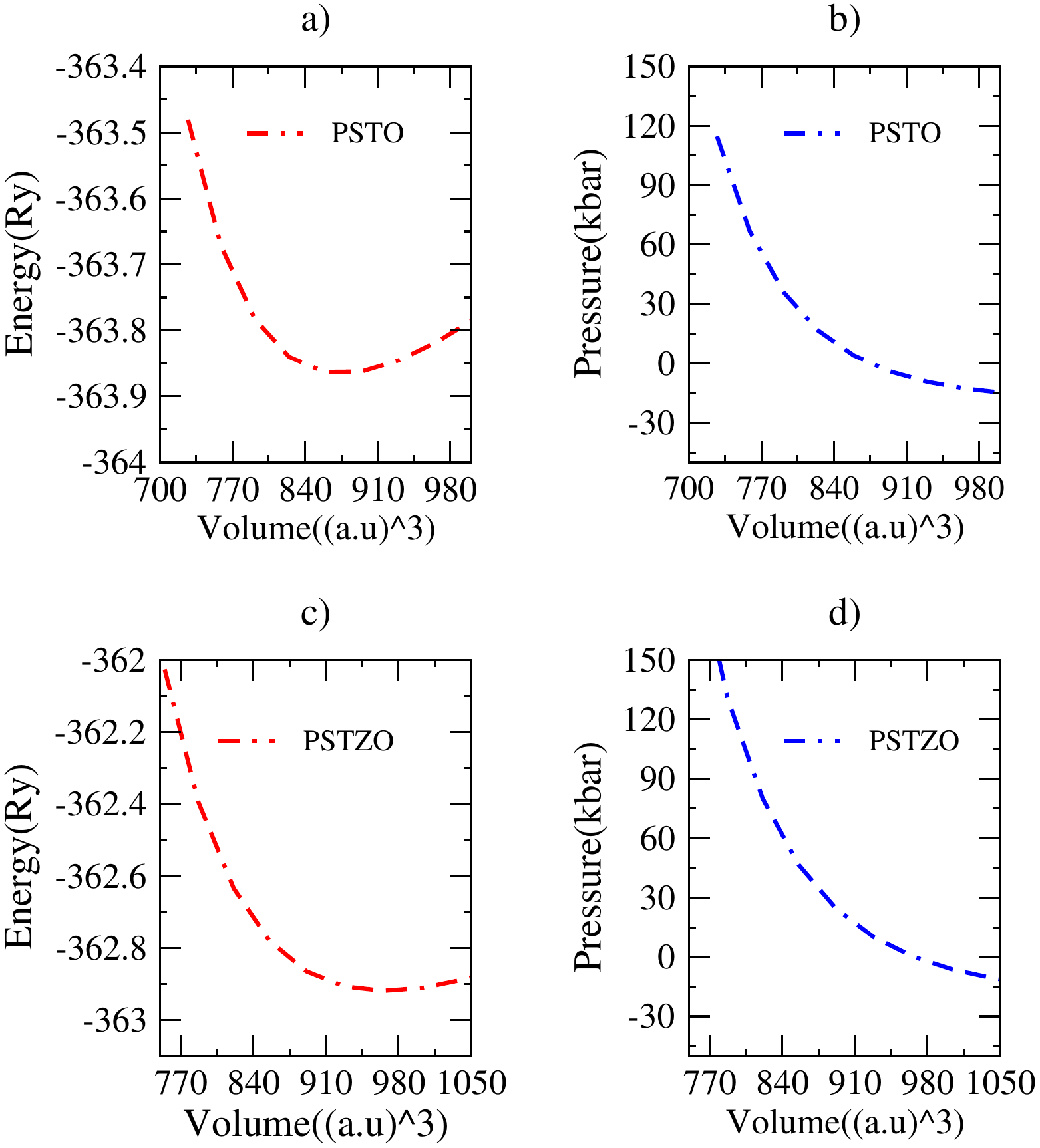}}
\caption{(Colour online) The total energy and pressure with respect to volume.} 
\label{figure3}
\end{figure}

 Moreover, equation of state (EOS) is a pressure-volume ($PV$) or energy-volume ($EV$) relation describing the behavior of a solid under compression or expansion. The relationship between $PV$ and $EV$ curve is shown in figure~\ref{figure3}. A series of total energy calculations as a function of volume can be fitted to an equation of state according to Murnaghan $E(V)$ curve \cite{murnaghan1944compressibility}.
    \begin{equation}
     E(V)=E_{o}+V\left\lbrace \frac{B}{B'}\bigg[\bigg(\frac{V_{o}}{V}\bigg)^{\frac{B'}{B'-1}}+1\bigg]-\frac{BV_{o}}{B'-1}\right\rbrace ,
    \end{equation}
where $E_{o}$, $V_{o}$, $B$, $B'$ are parameters of the fit: $E_{o}$ is the minimum energy, $V_{o}$ is the equilibrium cell volume, $B$ and $B'$ are the bulk modulus and its derivative of the material, respectively. The calculated values of lattice constant, bulk modulus, volume, and pressure derivatives of bulk modulus are summarized in table~\ref{table1}.
\begin{table}[!t]
\caption{ Computed equlibrium lattice constants, volume, bulk modulus and bulk modulus derivatives of compounds in tetragonal phase.}
\label{table1}
\vspace{1.5ex}
\begin{center}
 \begin{tabular}{|l|l|c|c|c|c|}
\hline 
  & Source &Lattice constants (\AA) & $V$(\AA)& $B$ & $B'$ \\ \hline \hline
Pb$_{0.5}$Sn$_{0.5}$TiO$_{3}$ & Our work  & $a=3.978$ and $c=8.353$  &129.59   & 129.59  & 10.39  \\ \hline
Pb$_{0.5}$Sn$_{0.5}$Ti$_{0.5}$(Zr$_{0.5}$)O$_{3}$& Our work  & $a=4.025$ and $c=8.616$  & 143.27 &200.2   & 9.70  \\ \hline
\end{tabular}
\end{center}
\end{table}
\subsection{Elastic properties}
Elastic constants are fundamental and indispensable to describe the mechanical properties of materials. They can be used to determine the elastic moduli, Poisson's ratio $n$, and elastic anisotropy factor of materials. Elastic properties are also closely associated with many fundamental solid-state properties, such as acoustic velocity, thermal conductivity, Debye temperature, inter-atomic potentials, and so on. When a crystal is subjected to an external loading (with the stress state described by a stress tensor $\sigma_{ij}$), its shape and dimensions are changed. The change of a crystal geometry can be described using the strain tensor ($\varepsilon_{kl}$). The relation between these two tensors (within the linear elastic limit) can be expressed by means of the generalized form of Hooke's law \cite{luan2018mechanical}.
\begin{equation}
 C_{ikl}=\bigg(\frac{\partial\sigma_{ij}}{\partial\varepsilon_{jkl}}\bigg)_{x} = \bigg(\frac{1}{V}\frac{\partial^{2}E}{\partial\varepsilon_{ij}\varepsilon_{kl}}\bigg)_{x}.
 \end{equation}
The nonzero components of the elastic constant tensor for each crystal point
group can be derived using the group theory \cite{nye1985physical}. The form of the elastic constants
tensor depends on the Laue class obtained by adding an inversion center to the
group operations. The Laue classes are eleven geometric crystal classes containing centrosymmetric crystallographic types of point groups and their subgroups \cite{nye1985physical}. Specifically, for tetragonal system the Laue class D4h (4/mmm) is used and there are also six independent elastic constants $C_{11}$, $C_{12}$, $C_{13}$, $C_{33}$, $C_{44}$ and $C_{66}$ which should satisfy the Born stability criteria \cite{weiner2012statistical}.
\[
 C_{11}-C_{12}>0,\quad C_{11}+C_{33}-2C_{13}>0,\quad C_{11}>0,
\]
\[
   C_{33}>0,\quad C_{44}>0,\quad C_{66}>0,
\]
\[
 2C_{11}+C_{33}+2C_{12}+4C_{13}>0
\]
and
\begin{equation}
\frac{1}{3}(C_{12}+2C_{13})<B<\frac{1}{3}(C_{11}+2C_{33}).
\end{equation}
We have calculated the six independent elastic constants of PSTO and PSTZO using stress-strain relation. At zero pressures, the calculated elastic constants of PSTO ($C_{11}=260.12$~GPa, $C_{12}=108.7$~GPa, $C_{13}=111.30$~GPa, $C_{33}=158.1$~GPa, $C_{44}=87.61$~GPa, and $C_{66}=94.1$~GPa). Similarly, the elastic constants of  PSTZO ($C_{11}=252.49$~GPa, $C_{12}=96.37$~GPa, $C_{13}=102.23$~GPa, $C_{33}=147.24$~GPa, $C_{44}=73.82$~GPa, and $C_{66}=79.11$~GPa). The computed elastic constants satisfy the mechanical stability conditions, and which are in reasonable agreement with those obtained by Marton and Elsasser \cite{marton2011first}. Moreover, the elastic modulus was calculated using the Voigt-Reuss-Hill approximation \cite{zuo1992elastic}. For a tetragonal system, Voigt bulk modulus ($B_\text{V}$) and shear modulus ($G_\text{V}$) are:
\begin{equation*}
 B_{v}=\frac{1}{9}[(C_{11}+C_{22}+C_{33})+2(C_{12}+C_{23}+C_{31})],
\end{equation*}
\begin{equation}
 G_{v}=\frac{1}{15}\left[ (C_{11}+C_{22}+C_{33})-(C_{12}+C_{23}+C_{31})+ 3(C_{44}+C_{55}+C_{66})\right] 
\end{equation}
and the Reuss bulk modulus ($B_\text{R}$) and shear modulus ($G_\text{R}$) are expressed as
\begin{equation*}
 B_\text{R}=[(S_{11}+S_{22}+S_{33})+2(S_{12}+S_{23}+S_{31})]^{-1},
\end{equation*}
\begin{equation}
 G_\text{R}=15[4(S_{11}+S_{22}+S_{33})-(S_{12}+S_{13}+S_{23})+
 3(S_{44}+S_{55}+_{66})]^{-1}
\end{equation}
the Hill$'$s average for the shear modulus ($G$) and bulk modulus ($B$) is given by
\begin{equation}
 G=\frac{1}{2}(G_{v}+G_\text{R}), \quad B=\frac{1}{2}(B_{v}+B_\text{R})
\end{equation}
while Young's modulus ($E$) and Poisson's ratio ($n$) are given by
\begin{equation}
 E=\frac{9BG}{3B+G}\,,\quad n=\frac{3B-2G}{2(3B+G)}.
\end{equation}
\begin{figure}[!t]
\centering
    {\includegraphics[width=0.42\textwidth]{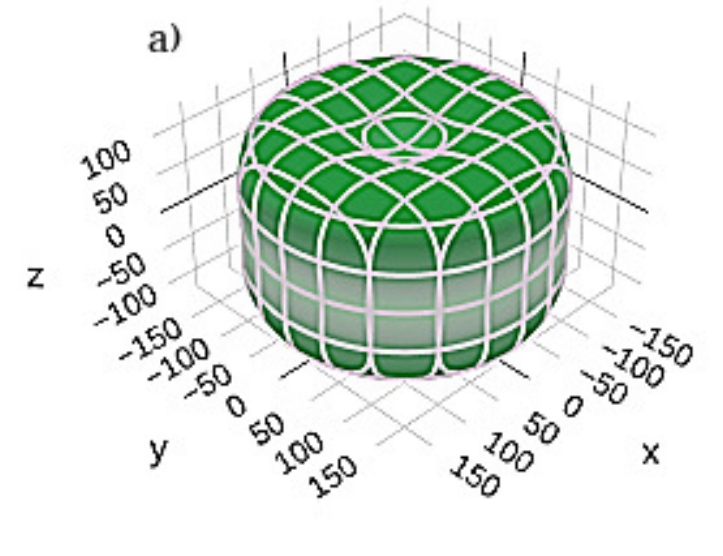}}
      \hspace{0.3cm}
    {\includegraphics[width=0.44\textwidth]{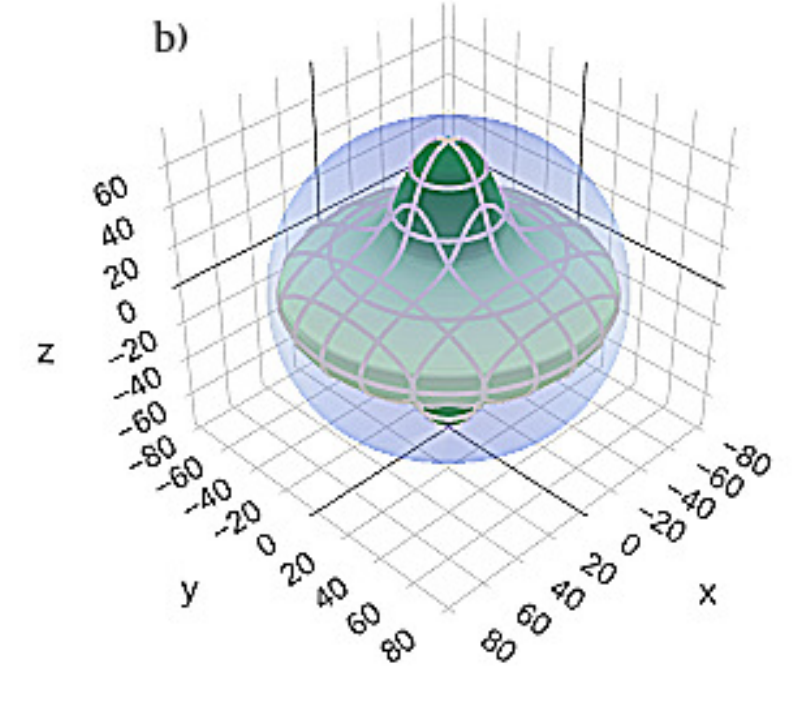}}
    \hspace{0.3cm}
    {\includegraphics[width=0.50\textwidth]{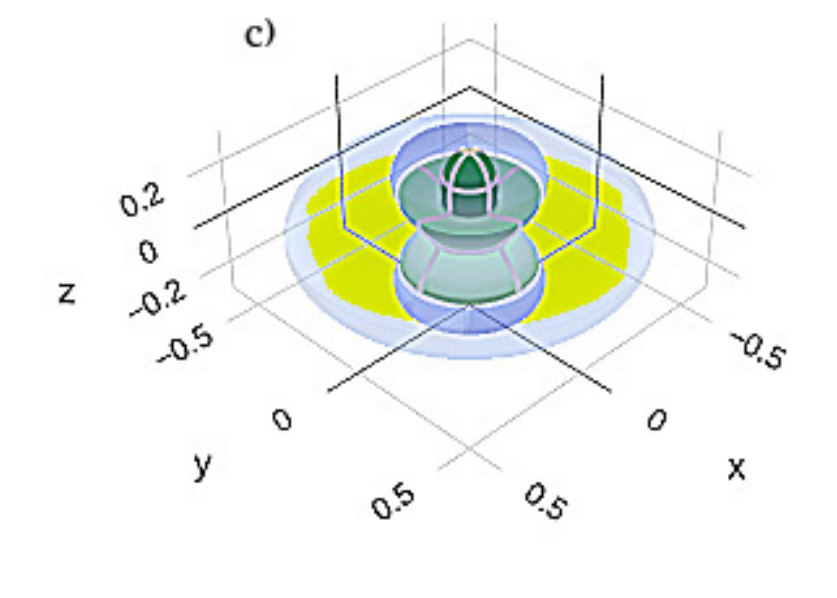}}
  \caption{(Colour online) Directional dependence of a) $E$, b) $G$ and c) $n$ of PSTZO. }
\label{figure4}
\end{figure}
\begin{figure}[!t]
\centering
    {\includegraphics[width=0.43\textwidth]{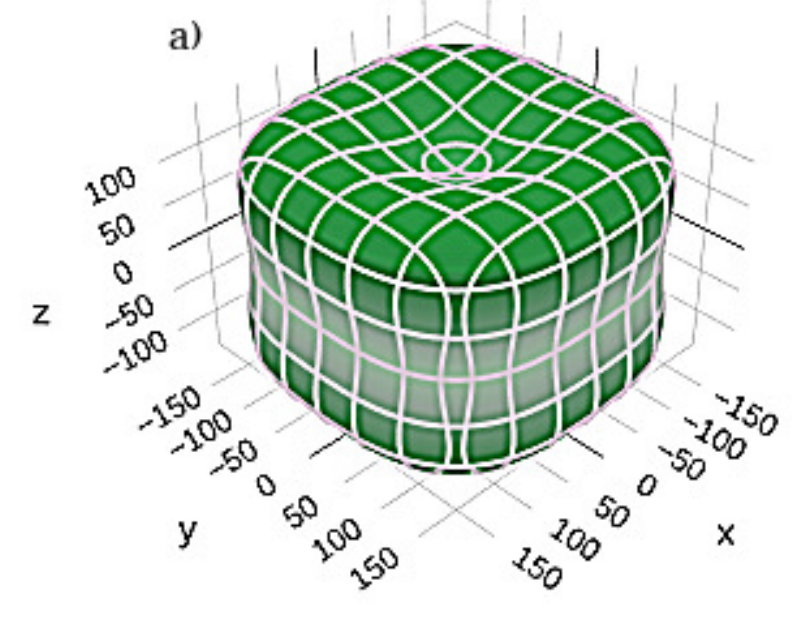}}
      \hspace{0.3cm}
    {\includegraphics[width=0.37\textwidth]{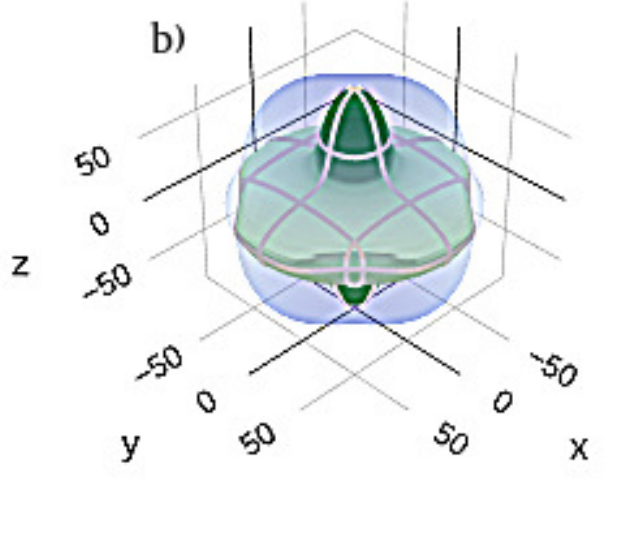}}
    \hspace{1cm}
    {\includegraphics[width=0.51\textwidth]{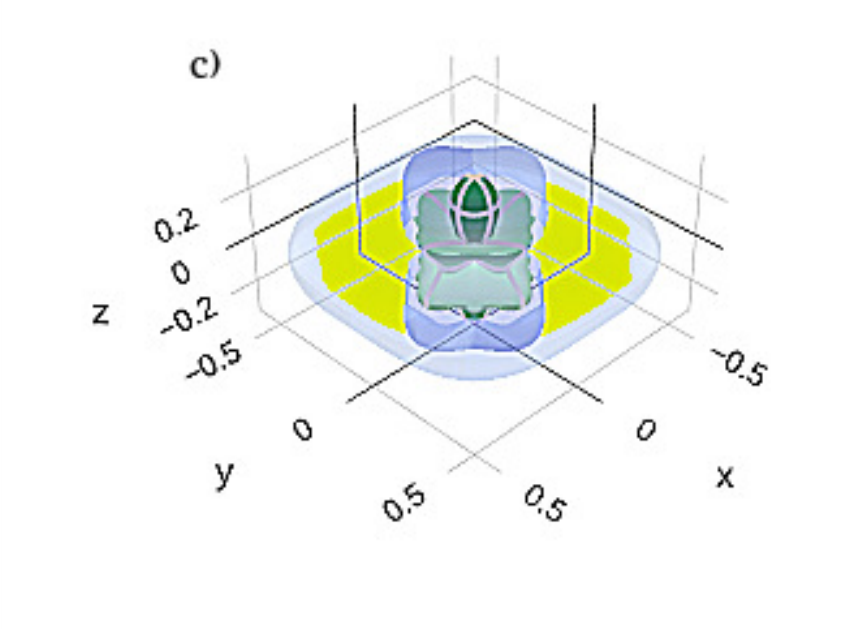}}
  \caption{(Colour online) Directional dependence of a) $E$, b) $G$ and c) $n$ of PSTO.}
  \label{figure5}
\end{figure}

The calculated bulk, shear, and Young's modulus of PSTO at zero pressure are 144.12, 72.45, and 186.15 GPa, respectively, which are in better agreement with the corresponding theoretical data \cite{marton2011first}. The obtained $B$, $G$, and $E$ of PSTZO are 134.34, 65.311, and 168.60 GPa, respectively, which are comparable with the theoretical values \cite{marton2011first}. From the results obtained, PSTO is stiffer than PSTZO. The $B/G$ ratio allows us to assess the ductility/brittleness of materials. According to Pugh \cite{pugh1954xcii}, 1.75 is the critical value that separates the brittleness and ductility behaviors of materials. When the ratio of $B/G$ is of a value higher than the critical one, then the material is associated with ductility. However, when the ratio of $B/G$ value is lower than the critical value, the material is considered to be brittle. Both PSTO ($B/G=1.989$) and PSTZO ($B/G=2.056$) compounds are categorized as ductile because the value of $B/G$ are higher than the critical value. The Poisson's ratio ($n$) is a mechanical parameter which provides useful information on the characteristic of the bonding forces. In the evaluation of Poisson's ratio, 0.25 and 0.5 are the lower and upper limits of the central force, respectively \cite{ravindran1998density}. The Poisson's ratio of 0.284 and 0.290 values are obtained for PSTO and PSTZO at zero pressure, respectively. This indicates that the inter-atomic forces are central. 

The universal anisotropic index ($A^{U}$) is a measure to quantify the elastic anisotropic characteristics based on the contributions of both bulk and sheared modulus \cite{ranganathan2008universal}:
\begin{equation}
\label{eq3.8}
 A^{U}=\frac{5G_{v}}{G_\text{R}}+\frac{B_{v}}{B_\text{R}}-6,
\end{equation}
where, $G_\text{V}$ and $B_\text{V}$ are sheared and bulk modulus obtained from Voigt approximation, respectively. Similarly, $G_\text{R}$ and $B_\text{R}$ are sheared and bulk modulus acquired from Reuss approximation. As it is described in the equation~(\ref{eq3.8}), when the universal anisotropic index ($A^{U}$) is equal to zero, the crystal is  isotropic. The variation from zero defines the level of elastic anisotropy. Therefore, the obtained universal anisotropic index at zero pressure ($A^{U}$) is 0.56 for PSTO and 0.70 for PSTZO.

For anisotropic materials, there is a real need to analyze and visualize the directional elastic properties such as Young's modulus, shear modulus and Poisson's ratio rather than their averages. Detailed tensorial analyses are also necessary to find materials with targeted or anomalous mechanical properties. We have calculated the orientation dependence of Young's modulus $E$, shear modulus $G$, and Poisson's ratio $n$ of both materials as depicted in figure \ref{figure4} and \ref{figure5}. As a result, the maximum and minimum values of $E$, $G$, and $n$ ratio along the normal direction of 111 and 100 planes  are summarized in table ~\ref{table2}. 

\begin{table}[!t]
\caption{Variations of the Young's modulus ($E$), shear modulus ($G$) and Poisson ratio ($n$).}
\label{table2}
\begin{center}
\begin{tabular}{|l|l|c|c|c|c|c|}
\hline
  & \multicolumn{2}{l|}{Young's modulus ($E$)} & \multicolumn{2}{l|}{shear modulus ($G$)} & \multicolumn{2}{l|}{Poisson ratio ($n$)} \\ \hline \hline
 &E$_\text{max}$&E$_\text{min}$&G$_\text{max}$&G$_\text{min}$&$n_\text{max}$&$n_\text{min}$\\ \hline
Pb$_{0.5}$Sn$_{0.5}$TiO$_{3}$  & 227.26 & 90.92& 94.11&42.92&0.66&0.06  \\ \hline
 Pb$_{0.5}$Sn$_{0.5}$Ti$_{0.5}$(Zr$_{0.5})$O$_{3}$ & 195.34 & 87.32& 79.11&42.04 &0.60&0.13  \\ \hline
\end{tabular}
\end{center}
\end{table}
As a fundamental parameter, the Debye temperature $\theta_\text{D}$ correlates with many physical properties of solids, such as specific heat, elastic constant and melting temperature. One of the standard methods to calculate the Debye temperature can be estimated from the averaged sound velocity $v_{m}$ is given by
\begin{equation}
 \theta_\text{D}=\frac{h}{k}\bigg[\frac{3n}{\piup}\bigg(\frac{N_\text{A}\rho}{M}\bigg)\bigg]^{\frac{1}{3}}v_{m}\,,
\end{equation}
where $h$ is the Planck's constant, $k$ is the Boltzmann's constant, $N_\text{A}$ is the Avogadro's number and $v_{m}$ is the average sound velocity. 
\begin{equation}
 v_{m}=\bigg[\frac{1}{3}\bigg(\frac{2}{v_{t}^{3}}+\frac{1}{v_{l}^{3}}\bigg)\bigg]^{-\frac{1}{3}},
\end{equation}
where, $v_{l}$ and $v_{t}$ are the longitudinal and transverse elastic sound velocities of the material that are determined by bulk modulus and shear modulus  \cite{long2013lattice, lu2014first}.
\begin{equation}
 v_{l}=\bigg(\frac{B+\frac{4}{3}G}{\rho}\bigg)^{\frac{1}{2}}
\end{equation}
\begin{equation}
 v_{t}=\bigg(\frac{G}{\rho}\bigg)^{\frac{1}{2}}.
\end{equation}
Based on the calculated elastic properties, the result of sound velocities and Debye temperatures  are listed in table~\ref{table3} accordingly. It can be pointed out that the calculated Debye temperature of PSTO is 466.451~K and PSTZO is 432.861~K.
\begin{table}[!t]
\caption{ The calculated  longitudinal, transverse, average sound velocity ($v_{l}$, $v_{t}$ and $v_{m}$ in m/s) and Debye temperatures ($\theta_\text{D}$ in~K).}
\label{table3}
\begin{center}
\vspace{2mm}
\begin{tabular}{|l|c|c|c|c|}
\hline
  & $v_{l}$ & $v_{t}$ & $v_{m}$ & $\theta_\text{D}$ \\ \hline \hline
Pb$_{0.5}$Sn$_{0.5}$TiO$_{3}$ &    6159.226 & 4765.584 & 3734.693 & 466.451   \\ \hline
Pb$_{0.5}$Sn$_{0.5}$Ti$_{0.5}$(Zr$_{0.5})$O$_{3}$ &   5624.279 & 4380.902 & 3381.050 & 432.861  \\ \hline
\end{tabular}
\end{center}
\end{table}
\subsection{Electronic properties}
Electronic band structures are a representation of the allowed electronic energy levels of solid materials and are used to better inform their electrical properties. 
\begin{figure}[!b]
\centering
{\includegraphics[width=0.75\textwidth]{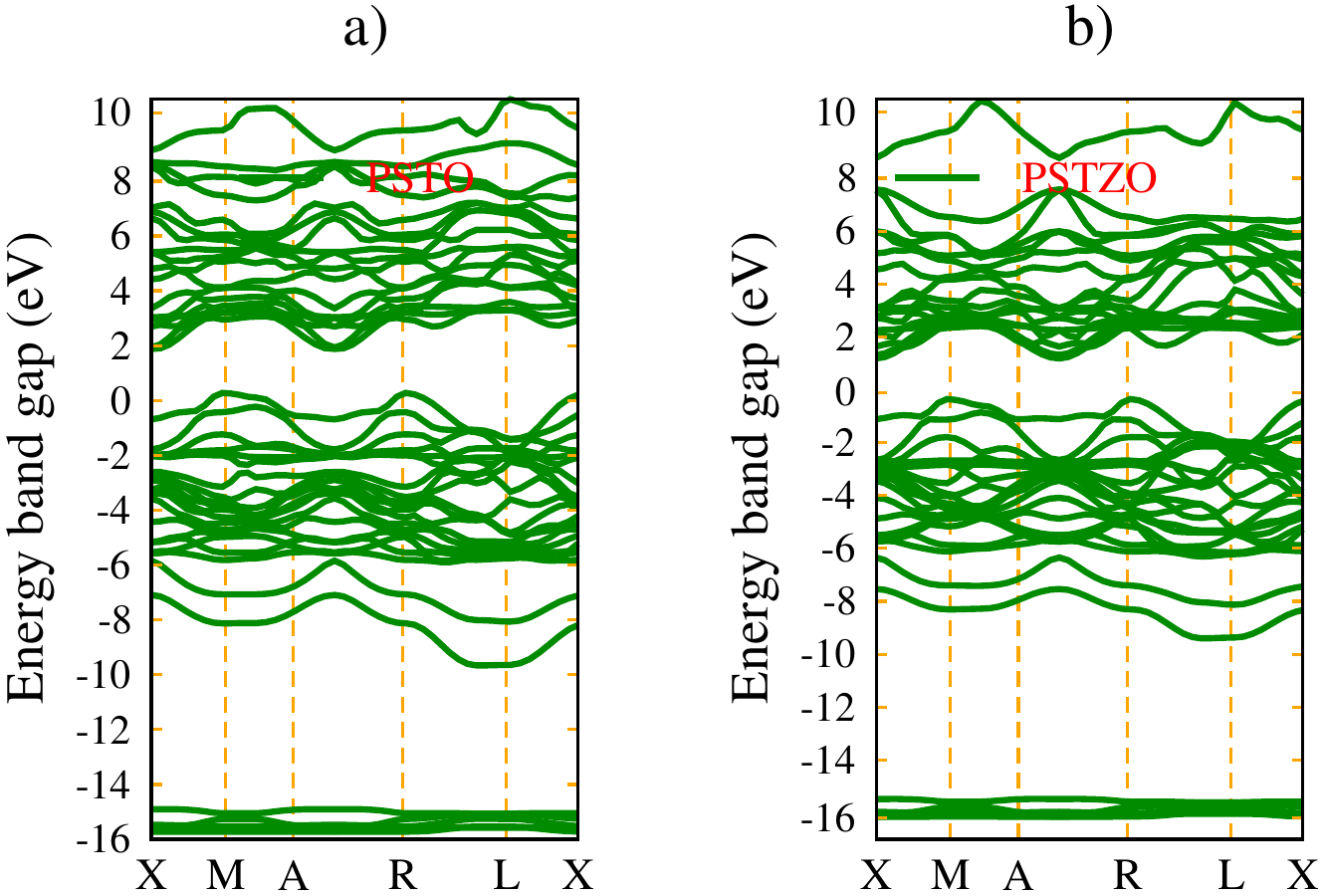}}
\caption{(Colour online) Electronic band structure of a) PSTO , and b) PSTZO.} 
\label{figure6}
\end{figure} 
From the calculations, the electronic band gap shows that the PSTO and PSTZO give the indirect band gap value of 1.51~eV, and 1.63~eV, respectively. The results obtained are in better agreement with the previous theoretical results~\cite{hussin2017theoretical}. The computed electronic band structure of the compounds is shown in figure~\ref{figure6} (a and b). Further explanation of the nature of the electronic band gap of ferroelectric compounds is carried out through the density of state. 
\begin{figure}[!t]
\centerline
{\includegraphics[width=0.92\textwidth]{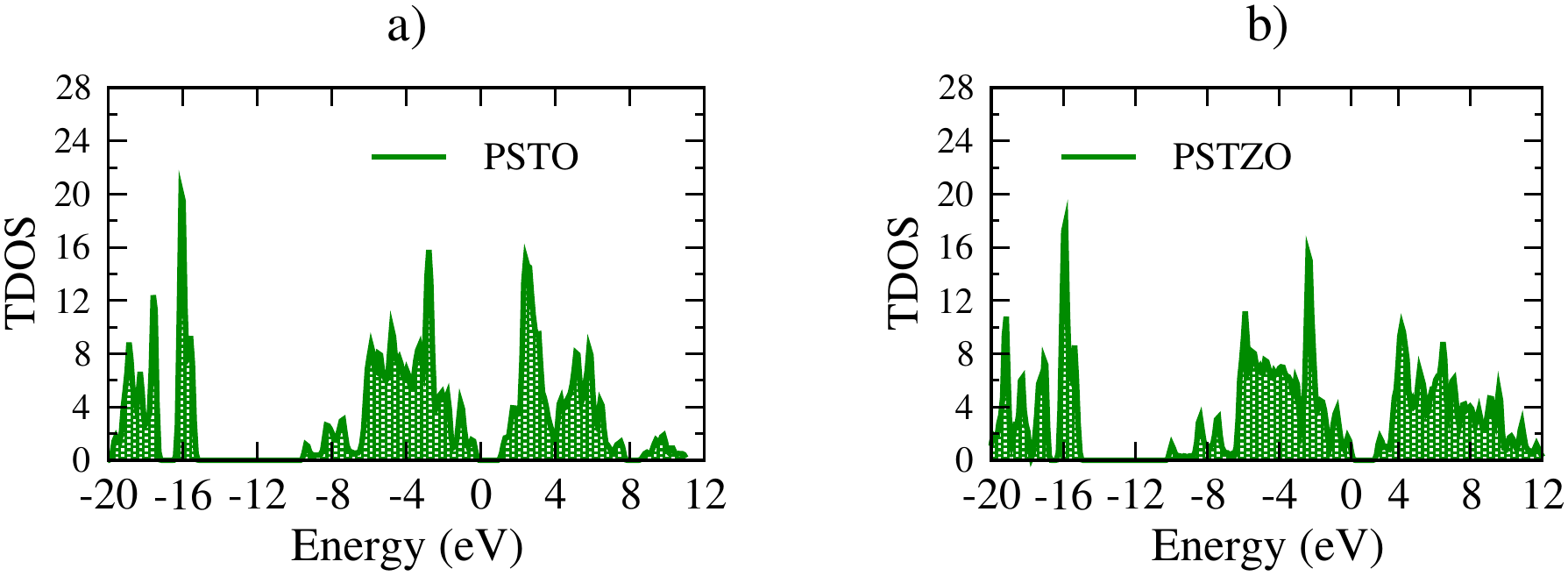}}
\caption{(Colour online) Total density of states (TDOS) of (a) PSTO, and b) PSTZO.} 
\label{figure7}
\end{figure}
In solid-state physics, density of states (DOS) shows the number of states per interval of energy as a function of energy, while the PDOS shows the projection of total DOS onto different angular momentum components. More specifically, from the PDOS, the contribution of specific atomic orbitals to the electronic bands and the interaction between different atomic orbitals is revealed. Total density of states (DOS) and projected density of states (PDOS) of tetragonal, PSTO and PSTZO are shown in figure~\ref{figure7} and~\ref{figure8}. The distribution of total density of states and partial density of states is plotted in the range of $-20.0$ to 12.0 eV.  The highest valence band of the compounds is mainly dominated by electron O 2$p$, Sn 5$s$ and Pb 6$s$ orbitals. While the lowest conduction band  mainly originates from the Ti 3$d$, Zr 4$d$, and Sn 5$p$ states. Thus, it can be inferred that both are a good ferroelectric material due to special hybridization between special lone pair Pb 6$s$ and O 2$p$ at valence band.
\begin{figure}[!b]
\centering
 a)
    {\includegraphics[width=0.35\textwidth]{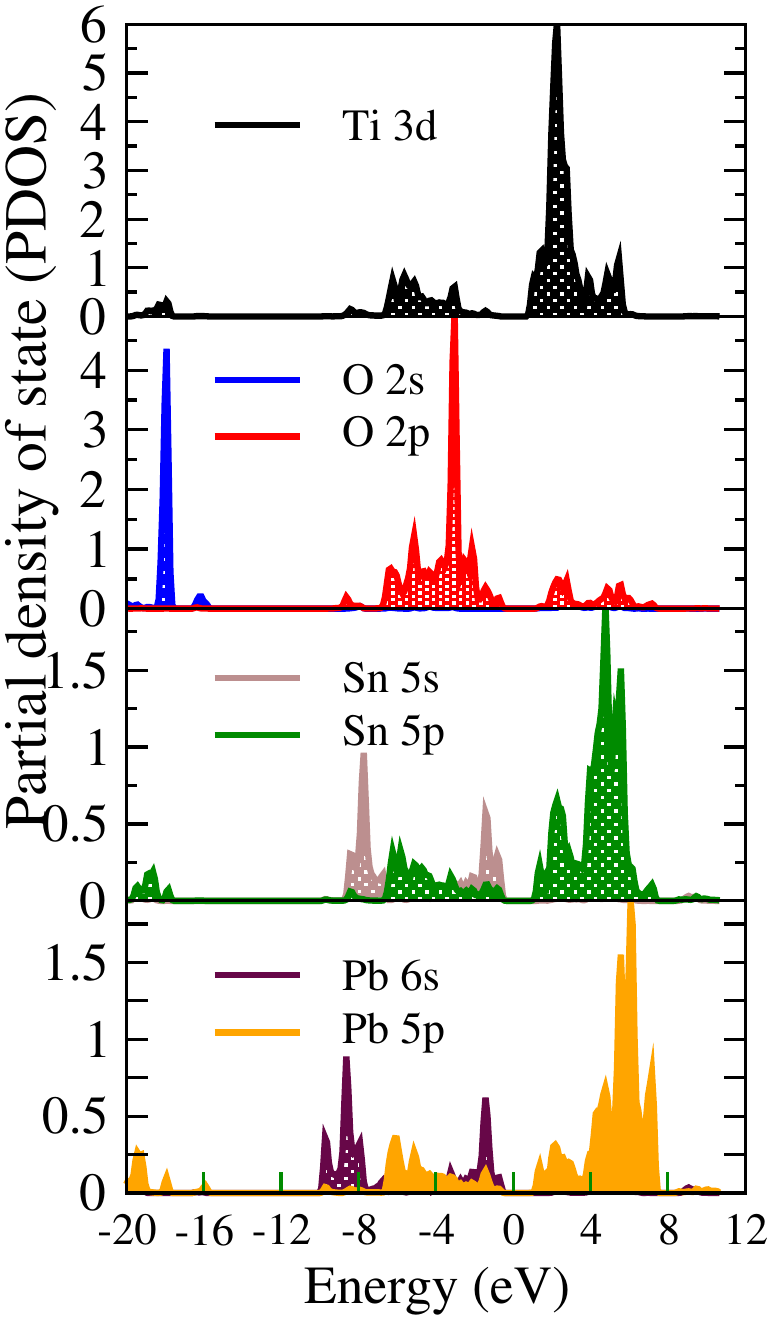}}
      \hspace{1cm}
  b) 
    {\includegraphics[width=0.36\textwidth]{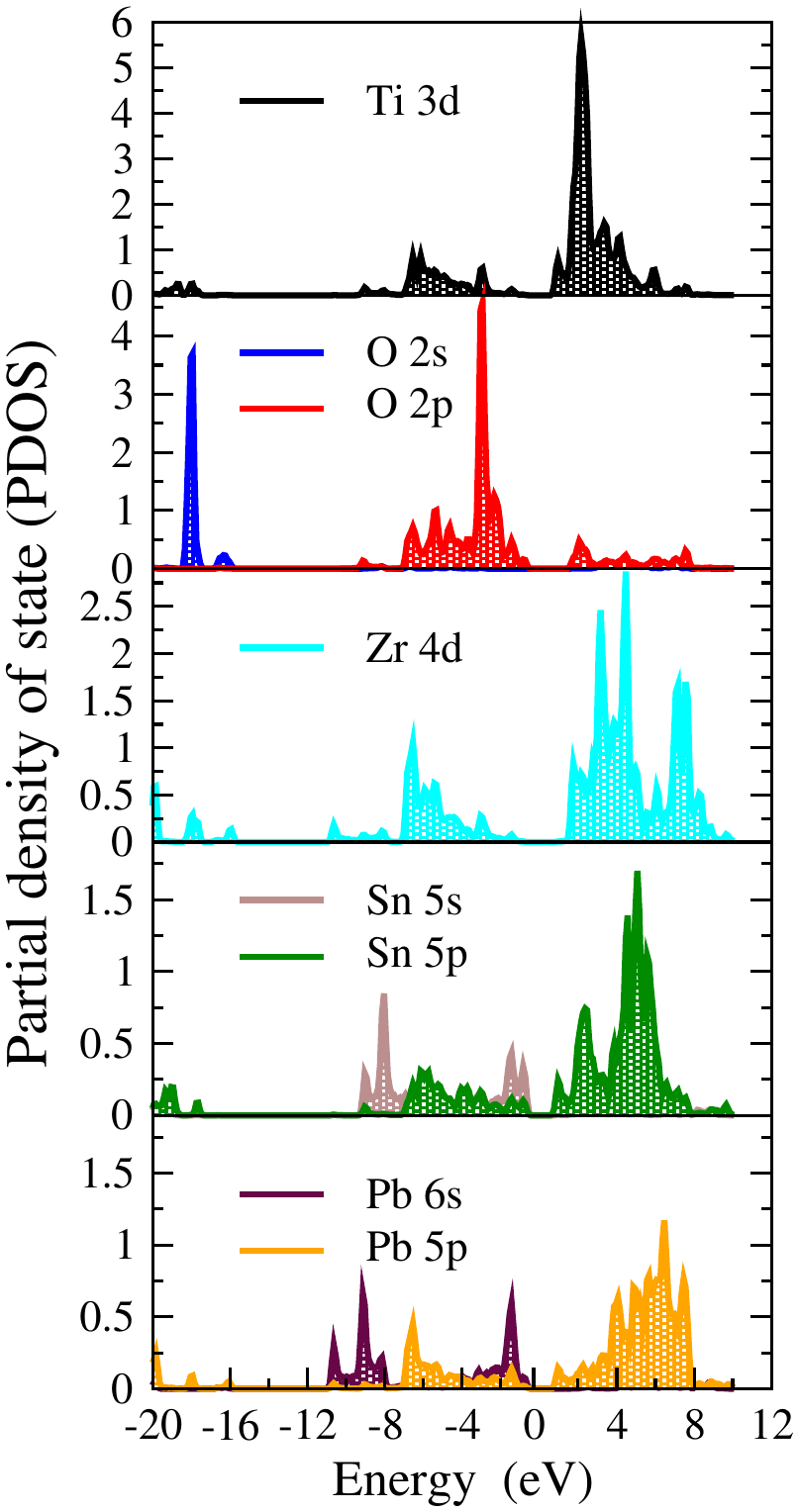}}
  \caption{(Colour online) Partial density of
states (PDOS) of a) PSTO and b) PSTZO.}
\label{figure8}
\end{figure}
\subsection{Born effective charges and spontaneous polarization}

\subsection*{Born effective charges}

The polarization induced by atomic displacements, given by the Born effective charges $Z^{*}$ \cite{gonze1992dielectric}, plays a key role in understanding both the polar ground state and the lattice dynamics. The Born effective charges are 
\begin{equation}
 Z^{*}_{i\alpha\beta} =\frac{\Omega}{e}\frac{\partial P_{\alpha}}{\partial u_{i\beta}},
\end{equation}
where $\alpha$ and $\beta$ denote directions. $P$ is the component of the polarization in the $\alpha^{th}$ direction and $u_{i\beta}$ is the periodic displacement of the $i^{th}$ atom in the $\beta^{th}$ direction. $\Omega$ is the volume and $e$ is the electron charge. Born effective charge (BEC or, $Z^{*}$), also known as transverse or dynamic effective charge, is a fundamental quantity that manifests coupling between lattice displacements and electrostatic fields \cite{gonze1992dielectric}. BEC is important in terms of theoretical study of ferroelectrics since the ferroelectric transition takes place from the competition of long-range coulomb interactions and short-range forces.
\begin{table}[!t]
\caption{ BEC tensor of Pb, Sn, Zr and Ti of Pb$_{0.5}$Sn$_{0.5}$TiO$_{3}$ and Sn$_{0.5}$Pb$_{0.5}$Ti$_{0.5}$(Zr$_{0.5}$)O$_{3}$ in ferroelectric phase.}
\vspace{1,5mm}
\label{table4}
\begin{center}
\begin{tabular}{|l|l|l|l||l|l|}
\hline
\multicolumn{3}{|l|}{\strut Pb$_{0.5}$Sn$_{0.5}$TiO$_{3}$}                                    & \multicolumn{3}{l|}{Sn$_{0.5}$Pb$_{0.5}$Ti$_{0.5}$(Zr$_{0.5}$)O$_{3}$}   \\ \hline \hline
\multirow{3}{*}{Z$^{*}_{Pb}$} &Z$^{*}$Pb$_{11}$ & 3.89705& \multirow{3}{*}{Z$^{*}_{Pb}$} &Z$^{*}$Pb$_{11}$ &3.91831\\ \cline{2-3} \cline{5-6} 
                   &     Z$^{*}$Pb$_{22}$ & 3.89705   &       & Z$^{*}$Pb$_{22}$ &3.91831  \\ \cline{2-3} \cline{5-6} 
                   &     Z$^{*}$Pb$_{33}$ & 3.79810   &       & Z$^{*}$Pb$_{33}$ & 2.41891 \\ \hline
\multirow{3}{*}{Z$^{*}_{Sn}$} & Z$^{*}$Sn$_{11}$ & 4.29538& \multirow{3}{*}{Z$^{*}_{Sn}$} &Z$^{*}$Sn$_{11}$&4.17773  \\ \cline{2-3} \cline{5-6} 
                   & Z$^{*}$Sn$_{22}$ & 4.29538&                &Z$^{*}$Sn$_{22}$& 4.17773  \\ \cline{2-3} \cline{5-6} 
                   & Z$^{*}$Sn$_{33}$ & 3.88430&                & Z$^{*}$Sn$_{33}$&2.15989   \\ \hline
\multirow{3}{*}{ Z$^{*}_{Ti}$} & Z$^{*}$Ti$_{11}$&6.98907& \multirow{3}{*}{Z$^{*}_{Ti}$} &Z$^{*}$Ti$_{11}$  &5.47072  \\ \cline{2-3} \cline{5-6} 
                   & Z$^{*}$Ti$_{22}$   &  6.98907 &            &Z$^{*}$Ti$_{22}$  & 5.47072 \\ \cline{2-3} \cline{5-6} 
                   & Z$^{*}$Ti$_{33}$   &  6.00741 &             &Z$^{*}$Ti$_{33}$  & 5.26044 \\ \hline
\multirow{3}{*}{}  & \multirow{3}{*}{} & \multirow{3}{*}{} & \multirow{3}{*}{Z$^{*}_{Zr}$}  &Z$^{*}$Zr$_{11}$&5.10754 \\ \cline{5-6} 
                   &                   &                   &      &Z$^{*}$Zr$_{22}$&5.10754  \\ \cline{5-6} 
                   &                   &                   &       &Z$^{*}$Zr$_{33}$&6.58694  \\ \hline
\end{tabular}
\end{center}
\end{table}

\begin{table}[!b]
\caption{BEC tensor of oxygen Pb$_{0.5}$Sn$_{0.5}$TiO$_{3}$ and Sn$_{0.5}$Pb$_{0.5}$Ti$_{0.5}$(Zr$_{0.5}$)O$_{3}$ in ferroelectric phase.}
\label{table5}
\small
\begin{center}
\begin{tabular}{|c|ccc|ccc|ccc|}
\hline
  & \multicolumn{3}{|c|}{O$_{1}$} & \multicolumn{3}{|c|}{O$_{2}$} & \multicolumn{3}{|c|}{O$_{3}$} \\ \hline \hline
\multirow{3}{*}{Pb$_{0.5}$Sn$_{0.5}$TiO$_{3}$} &$ -2.76$&0&0 & $ -5.74$&0&0  &$ -2.44$&0&0\\
\multirow{3}{*}{}  & 0&$-2.76$&0 & 0&$-2.44$&0& 0&$-5.74$&0\\
\multirow{3}{*}{}  & 0&0&$-4.94$& 0&0& $-2.51$& 0&0&$-5.51$\\ \hline 
\multirow{3}{*}{Sn$_{0.5}$Pb$_{0.5}$Ti$_{0.5}$(Zr$_{0.5}$)O$_{3}$} & $-1.92$&0&0 &  $-4.51$&0&0  & $-2.80$&0&0\\
\multirow{3}{*}{}  & 0&$-1.92$&0 & 0&$-2.80$&0 &0&$-4.51$&0\\
\multirow{3}{*}{}  & 0&0&$-4.38$& 0&0&$-1.81$& 0&0&$-1.81$\\ \hline 
\end{tabular}
\end{center}
\end{table}
In the present work, we have evaluated the BEC tensors of each ion in tetragonal phase of PSTO and PSTZO in the framework of density functional perturbation theory (DFPT). As presented in tables~\ref{table4} and~\ref{table5}, the Born effective dynamical charge of each compounds atom is larger than the nominal ionic charge. The calculated BECs are in agreement with the former results \cite{saghi1999first}. The large values of BEC of each atom compared to the nominal ionic charge show the importance of the ions as the driving force of the ferroelectric distortion.
\subsection*{Spontaneous polarization}
An accurate quantitative method for computing the polarization to all orders in displacement is the so-called Berry phase (or modern) theory of polarization \cite{resta2007theory}. The spontaneous polarization (P) of the compounds is obtained from the sum of both ionic polarization ($P_\text{ion}$) and electronic polarization ($P_\text{el}$) 
\begin{equation}
 P_\text{ion}+P_\text{el}=\frac{|e|}{\varOmega}\sum_{k}Z_{k}u_{k}+\bigg(-\frac{2|e|\ri}{2(\piup)^{3}}\int_{A}\rd k_{\bot }\sum_{n=1}^{M}\int_{0}^{G_{\parallel }}\langle u_{nk}|\frac{\partial}{\partial k_{\parallel }}| u_{nk} \rangle \rd k_{\parallel } \bigg ), 
\end{equation}
where in the electronic contribution, the sum $n$ runs over all $M$ occupied bands, and where $k_{\parallel }$ is parallel to the direction of polarization, and $G_{\parallel }$ is a reciprocal lattice vector in the same direction. The states $u_{nk}$ is the lattice-periodical part of the Bloch wave function. The integral over the perpendicular directions can easily be converged with a few  k-points. The ionic part of polarization is a well-defined quantity from electromagnetic theory. However, the electronic part of of polarization cannot be directly evaluated on the basis of localized contributions. The calculated spontaneous polarization at the equilibrium lattice constant using Berry's phase approach was determined to be 0.8662~C/m$^{2}$ (PSTO) and 1.0824~C/m$^{2}$ (PSTZO). The computed spontaneous polarization shows a high ferroelectric behavior compared to prototypical ferroelectric perovskites~\cite{zhang2017comparative, kuma2019structural}.
\subsection{Optical dielectric functions}
\begin{figure}[!t]
\centering
{\includegraphics[width=0.75\textwidth]{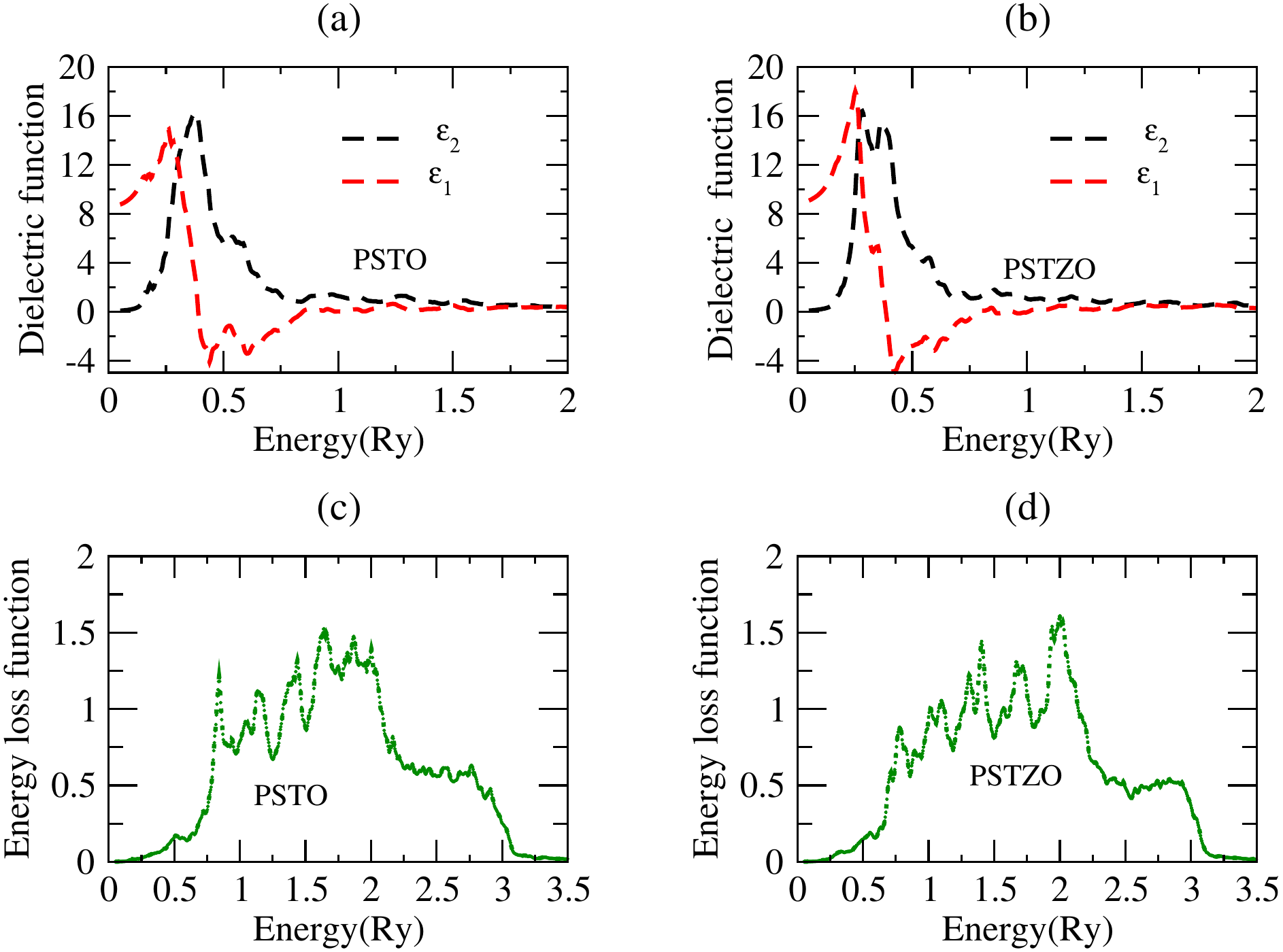}}
\caption{(Colour online) Calculated real ($\varepsilon_{1}(\omega)$), imaginary ($\varepsilon_{2}(\omega)$) dielectric function and energy-loss function L($\omega$). } \label{figure9}
\end{figure}

The optical properties can be explained in detail through the knowledge of the complex dielectric function $\varepsilon(\omega)=\varepsilon_{1}(\omega)+\ri\varepsilon_{2}(\omega)$. The imaginary part of the dielectric function $\varepsilon_{2}(\omega)$ is calculated by the sum of all possible direct transitions from the occupied to unoccupied states over the Brillouin zone \cite{gajdovs2006linear}.
\begin{equation}
    \varepsilon_{2}(\omega)=\frac{2\piup e^{2}}{\Omega\varepsilon_{o}}\sum_{k,v,c}|\langle
    \psi^{c}_{k}|^{u}r|\psi^{v}_{k}\rangle|^{2}\delta[E_{k}^{c}-E_{k}^{v}-E],
\end{equation}
where $e$ refers to electron charge, and $\psi^{c}_{k}$ and $\psi^{v}_{k}$ are the conduction band (CB) and valence band (VB) wave functions at $k$, respectively. The real and imaginary dielectric functions are linked by the Kramer-Kronig relation, which is used to calculate the real part $\varepsilon_{1}(\omega)$ of the dielectric function
\begin{equation}
    \varepsilon_{1}(\omega)=1+ \frac{2}{\piup}P\int_{0}^{\infty}\frac{\omega'\varepsilon_{2}(\omega')}{\omega'^{2}-\omega^{2}}\rd\omega',
\end{equation}
where P is the principal value of the integral,
 $P=\lim_{a\to 0} \int_{-\infty}^{\omega-a}\frac{\varepsilon(\omega)}{\omega' - \omega}\rd\omega'+ \lim_{a \to 0}\int_{\omega+a}^{+\infty}\frac{\varepsilon(\omega')}{\omega'-\omega}\rd\omega'$.
 As it is shown in figure \ref{figure9} (a and b), the real and the imaginary parts of complex dielectric function completely describe the optical properties of a medium for different photon energies. For the real part $\varepsilon_{1}(\omega)$ of the dielectric function, the highest peak for PSTO it appears around 0.23~Ry (3.13~eV),  while for PSTZO it appears at 0.25~Ry (3.40~eV). While, the imaginary part $\varepsilon_{2}(\omega)$ of the dielectric function illustrates the optical transition mechanism. 
Each peak in the imaginary part of the dielectric function corresponds to an electronic transition. The major $\varepsilon_{2}(\omega)$ peaks for PSTO located at 0.37~Ry (5.03~eV), 0.54~Ry (7.34~eV), 1.27~Ry (17.27~eV) and 1.51~Ry (20.54~eV), correspond to the transition from occupied VB O 2$p$, Pb 6$s$, Sn 5$s$ and O 2$s$ to unoccupied CB Ti 3$d$, Pb 5$p$ and Sn 5$p$. For PSTZO, the $\varepsilon_{2}(\omega)$ peaks 0.28~Ry (3.80~eV), 0.36~Ry (4.89~eV), 0.85~Ry (11.56~eV) and 1.89~Ry (25.71~eV) originate from the transition of O 2$p$, Pb 6$s$, Sn 5$s$ and O 2$s$ occupied to unoccupied CB Ti 3$d$, Zr 4$d$, Pb 5$p$ and Sn 5$p$. In the light of $\varepsilon_{1}(\omega)$ and $\varepsilon_{2}(\omega)$, the electron energy-loss function is defined as 
\begin{equation}
 L(\omega) =-\Im\bigg(\frac{1}{\varepsilon(\omega)}\bigg)=\frac{\varepsilon_{2}(\omega)}{\varepsilon_{1}^{2}(\omega)+ \varepsilon_{2}^{2}(\omega)}.
\end{equation}
The electron energy loss function $L(\omega)$ is another useful tool to investigate the behavior of a material with the light. This property of a medium or a material measures the propagation loss of energy inside the medium or material.  As shown in  figure \ref{figure9} (c and d) the major peak of $L(\omega)$ occurs at 1.76~Ry (23.94~eV) for PSTO and at 2.1~Ry (28.57~eV) for PSTZO. The sharp peaks of $L(\omega)$ associated with the plasma oscillation and their corresponding frequencies are called plasma frequencies \cite{he2010first}.

\subsection{Conclusions}
In summary, the structural, elastic, electronic and optical properties of PSTO and PSTZO were extensively studied using pseudo-potential plane-wave (PPs-PW) approach in the framework of density functional theory. The exchange-correlation potential was calculated within the framework of generalized gradient approximation (GGA). The equilibrium lattice parameter, bulk modulus and its pressure derivative were computed from equation of state (EOS). The obtained structural lattice constant shows a better agreement with the previous theoretical data. From the study of elastic properties, the calculated bulk modulus $B$, Young's modulus $E$, shear modulus $G$ and Poisson's ratio $n$ of PSTO are 144.12, 72.45, 186.15~GPa and 0.284, respectively. The obtained $B$, $G$, $E$ and $n$ of PSTZO are also 134.34, 65.311, 168.60~GPa and 0.29, respectively, which are consistent with the previously reported data. The computed $B/G$ ratio implies that both systems are ductile materials. Moreover, regarding the band structure, we found that the PSTO compound has 1.51~eV indirect bandgap, whereas PSTZO is found to be 1.63~eV indirect bandgap. The optical properties such as dielectric constant, and energy loss function were also investigated.

 The calculated BEC of atoms of both compounds  are larger than the nominal ionic charge, which are in agreement with the former results. The large values of Born effective dynamical charge show the importance of the ions as the driving force of the ferroelectric distortion. Using the Berry phase approach, spontaneous polarization was determined to be 0.8662~C/m$^{2}$ (PSTO) and 1.0824~C/m$^{2}$ (PSTZO). The obtained spontaneous polarization of both materials shows a high ferroelectric behavior compared to prototypical ferroelectric materials and they can be the best choice of future ferroelectric materials. 
\section*{Acknowledgements}
The author would like to acknowledge the East African Institute for fundamental Research (EAIFR), ICTP's partner institute for High Performance Clusters machine support during the computation.  

\newpage
\ukrainianpart
\title{Першопринципні дослідження структурних, пружних, електронних та оптичних властивостей  Pb$_{0.5}$Sn$_{0.5}$TiO$_{3}$ і Pb$_{0.5}$Sn$_{0.5}$Ti$_{0.5}$(Zr$_{0.5}$)O$_{3}$} 

\author{Ш.Г. Кума\refaddr{label1},
     M.M. Волдемаріам \refaddr{label2}}
\addresses{
\addr{label1} Фізичний факультет,Університет Воллеги, P.O.Box 395, Некемте, Ефіопія
\addr{label2} Фізичний факультет, Університет м. Джімма, P.O.Box 378, Джімма, Ефіопія}
\makeukrtitle
\begin{abstract}
Структурні, електронні, пружні та оптичні властивості тетрагональної (P4мм) фази Pb $ _ {0,5} $ Sn $ _ {0,5} $ TiO $ _ {3} $ (PSTO) та Pb $ _ {0,5} $ Sn $ _ {0,5} $ Ti $ _ {0,5} $ (Zr $ _ {0,5} $) O $ _ {3} $ (PSTZO) досліджуються за допомогою першопринципних розрахунків  на основі теорії функціоналу густини (DFT) з використанням процедури псевдопотенціальної плоскої хвилі (PP-PW)  в рамках узагальненого градієнтного наближення  (GGA). Ми розрахували такі властивості основного стану, як  рівноважні сталі  гратки, об’єм, об’ємний модуль та  похідну від нього по тиску. З пружних констант механічні параметри, такі як коефіцієнт анізотропії, модуль пружності та коефіцієнт Пуассона, отримуються із середнього наближення Фойгта-Ройсса-Хілла. Замість їхніх середніх значень,  залежність модуля пружності та коефіцієнта Пуассона за напрямками моделюються та візуалізуються з урахуванням пружних властивостей обох систем. Крім того, отримуються деякі нові результати, такі як температура Дебая та швидкість звуку. Більше того, ми представили результати електронної зонної структури, густини станів та густини заряду. Ці результати добре узгоджуються  з існуючими теоретичними даними. Також обчислено оптичну діелектричну функцію та спектр втрат енергії обох систем. Ефективний заряд (BEC) кожного атома для обох систем обчислюється на основі теорії збурення функціоналу густини (DFPT). Нарешті, спонтанна поляризація також визначається із сучасної теорії поляризації: 0,8662 C/м$^{2}$ (PSTO) та 1,0824~C/м$^{2}$ (PSTZO). 	

\keywords DFT,  PSTO і PSTZO, електронні, пружні та оптичні властивості
\end{abstract}

\lastpage

\begin{thebibliography}{99}
\bibitem{rodel2009perspective}
   Rodel J., Jo W., Seifert K.T., Anton E.M., Granzow T., Damjanovic D., J. Am. Ceram. Soc., 2009, 
\\    \textbf{92},
    1153--1177, \doi{10.1111/j.1551-2916.2009.03061.x}.
\bibitem{kittel1969einfuhrung} 
    Kittel C., Gress J.M., Lessard A., Einf\"{u}hrung in die Festk\"{o}rperphysik, 1969.
\bibitem{dove1997theory} 
    Dove M.T., Am. Mineral., 1997, 
    \textbf{82},
    213--244, \doi{10.2138/am-1997-3-401}.
\bibitem{resta1994macroscopic} 
    Resta R., Rev. Mod. Phys., 1994, 
    \textbf{66},
    899, \doi{10.1103/RevModPhys.66.899}.
\bibitem{knapp2006site} 
    Knapp M.C., Woodward P.M., J. Solid State Chem., 2006, 
    \textbf{179},
    1076--1085, \\\doi{10.1016/j.jssc.2006.01.005}.
\bibitem{jaffe2012piezoelectric} 
    Jaffe B., Piezoelectric Ceramics, Elsevier, Amsterdam, 2012.
\bibitem{sani2004high} 
    Sani A., Noheda B., Kornev I.A., Bellaiche L., Bouvier P., Kreisel J., Phys. Rev. B, 2004,
    \textbf{69},
    020105, \\\doi{10.1103/PhysRevB.69.020105}.
\bibitem{noheda2000stability} 
    Noheda B., Cox D.E., Shirane G., Guo R., Jones B., Cross L.E., Phys. Rev. B, 2000, 
    \textbf{63},
    014103, \\\doi{10.1103/PhysRevB.63.014103}.
\bibitem{noheda2002phase} 
    Noheda B., Cox D.E., Shirane G., Gao J., Ye Z.G., Phys. Rev. B, 2002, \textbf{66}, 
    054104,\\ \doi{10.1103/PhysRevB.66.054104}.
\bibitem{wang2014first} 
    Wang L., Yuan P., Wang F., Liang E., Sun Q., Guo Z.,  Jia Y., Bull. Mater. Sci., 2014, 
    \textbf{49},
    509--513,\\ \doi{10.1016/j.materresbull.2013.08.075}.
\bibitem{bell2018lead}
    Bell A.J., Deubzer O., MRS Bulletin, 2018, 
    \textbf{43},
    581--587, \doi{10.1557/mrs.2018.154}.
\bibitem{robinson2009waste} 
    Robinson B.H., Sci. Total Environ., 2009, \textbf{408},
    183--191, \doi{10.1016/j.scitotenv.2009.09.044}.
\bibitem{babu2007electrical} 
    Babu B.R., Parande A.K., Basha C.A.,
    Waste Manag. Res., 2007, \textbf{25}, 307--318,\\ \doi{10.1177/0734242X07076941}.
\bibitem{taib2014first} 
    Taib M.F.M., Yaakob M.K., Badrudin F.W., Rasiman M.S.A., Kudin T.T., Hassan O.H., Yahya M.Z.A., \\Integr. Ferroelectr., 2014, 
    \textbf{155},
    23--32, \doi{10.1080/10584587.2014.905105}.
\bibitem{zhao2017lead}
    Zhao L., Liu Q., Gao J., Zhang S., Li J.F., Adv. Mater.,  2017, 
    \textbf{29},
    1701824, \doi{10.1002/adma.201701824}.
\bibitem{coondoo2013lead}
    Coondoo I., Panwar N., Kholkin A., Adv. Dielectr., 2013, 
    \textbf{3},
    1330002, \doi{10.1142/S2010135X13300028}.
\bibitem{cohen2011class} 
    Cohen R., Ganesh P., Class of Pure Piezoelectric Materials, United State Patent: US 2009/0291324 A1.
\bibitem{roy2011theory} 
    Roy A., Vanderbilt D., Phys. Rev. B, 2011, 
    \textbf{83},
    134116, \doi{10.1103/PhysRevB.83.134116}.
\bibitem{giannozzi2009quantum} 
   Giannozzi P., Baroni S., Bonini N., Calandra M., Car R., Cavazzoni C., Ceresoli D., Chiarotti G.L., \\Cococcioni M., Dabo I., Dal Corso A., J. Phys.: Condens. Matter, 2009, 
\textbf{21},
395502, \\\doi{10.1088/0953-8984/21/39/395502}.
\bibitem{perdew1996generalized} 
    Perdew J.P., Burke K., Ernzerhof M.,  Phys. Rev. Lett., 1996, \textbf{77},
    3865, \doi{10.1103/PhysRevLett.77.3865}.
\bibitem{hasnip2006electronic}
    Hasnip P.J., Pickard C.J., Comput. Phys. Commun., 2006,
    \textbf{174},
    24--29, \doi{10.1016/j.cpc.2005.07.011}.
\bibitem{monkhorst1976special}
    Monkhorst H.J., Pack J.D., Phys., Rev. B, 1976,
    \textbf{13}, 
    5188, \doi{10.1103/PhysRevB.13.5188}.
\bibitem{gonze1997dynamical}
    Gonze X., Lee C., Phys. Rev. B, 1997,
    \textbf{55},
    10355, \doi{10.1103/PhysRevB.55.10355}.
\bibitem{resta2007theory}
    Resta R., Vanderbilt D., In: Physics of Ferroelectrics. Topics in Applied Physics, Vol 105. Springer, Berlin, Heidelberg, 2007, 31--68, 
    \doi{10.1007/978-3-540-34591-6_2}.
\bibitem{hussin2017theoretical}
    Hussin N.H., Taib M.F.M., Hassan O.H., Yahya M.Z.A.,  Mater. Res. Express, 2017,
    \textbf{4}, 
    074001, \\\doi{10.1088/2053-1591/aa6c42}.
\bibitem{wang2010synthesis}
    Wang J., Pang X., Akinc M., Lin Z., J. Mater. Chem., 2010, \textbf{20}, 
    5945--5949, \doi{10.1039/C0JM00270D}.
\bibitem{murnaghan1944compressibility}
   Murnaghan F.D., PNAS, 1944, 
    \textbf{30},
    244, \doi{10.1073/pnas.30.9.244}.
\bibitem{luan2018mechanical}
    Luan X., Qin H., Liu F., Dai Z., Yi Y., Li Q., Crystals, 2018, 
    \textbf{8}, 
    307, \doi{10.3390/cryst8080307}.
\bibitem{nye1985physical}
    Nye J.F., Physical Properties of Crystals: Their Representation by
    Tensors Matrices, \\Oxford University Press, Oxford, 1985.
\bibitem{weiner2012statistical}
    Weiner J.H., Statistical Mechanics of Elasticity, Dover Publications, New York, 2012.
\bibitem{marton2011first}
    Marton P., Els\"{a}sser C., Phys. Status Solidi B, 2011,
\textbf{ 248},
    2222--2228, \doi{10.1002/pssb.201046598}.
\bibitem{zuo1992elastic}
    Zuo L., Humbert M., Esling C., J. Appl. Cryst.,
    1992, 
    \textbf{25},
    751--755, \doi{10.1107/S0021889892004874}.
\bibitem{pugh1954xcii}
    Pugh S.F., Philos. Mag., 1954,
    \textbf{ 45},
    823--843, \doi{10.1080/14786440808520496}.
\bibitem{ravindran1998density}
    Ravindran P., Fast L., Korzhavyi P.A., Johansson B., Wills J., Eriksson O., J. Appl. Phys.,\\ 1998, 
   \textbf{84},
   4891--4904, \doi{10.1063/1.368733}.
\bibitem{ranganathan2008universal}
    Ranganathan S.I., Ostoja-Starzewski M., Phys. Rev. Lett., 2008, \textbf{101}, 
   055504,\\ \doi{10.1103/PhysRevLett.101.055504}.
\bibitem{long2013lattice}
    Long J., Yang L., Wei X., J. Alloys Compd., 2013, \textbf{549}, 336--340, \doi{10.1016/j.jallcom.2012.08.120}.
  
  \bibitem{lu2014first}
        Lu H., Long J., Yang L., Huang W., Int. J. Mod. Phys. B, 2014, \textbf{28}, 1450057,
        \doi{10.1142/S021797921450057X}.
   
\bibitem{gonze1992dielectric}
    Gonze X., Allan D.C., Teter M.P., Phys. Rev. Lett., 1992, 
    \textbf{68}, 3603, \doi{10.1103/PhysRevLett.68.3603}.
\bibitem{saghi1999first}
    Saghi-Szabo G., Cohen R.E., Krakauer H., Phys. Rev. B, 1999, \textbf{59}, 12771, \doi{10.1103/PhysRevB.59.12771}.
\bibitem{zhang2017comparative}
    Zhang Y., Sun J., Perdew J.P., Wu X., Phys. Rev. B, 2017, \textbf{96}, 
    035143, \doi{10.1103/PhysRevB.96.035143}.
\bibitem{kuma2019structural}
    Kuma S., Woldemariam M.M., Adv. Condens. Matter Phys., 
    \textbf{2019},
    2019, \doi{10.1155/2019/3176148}.
\bibitem{gajdovs2006linear}
   Gajdo\v{s} M., Hummer K., Kresse G., Furthm\"{u}ller J., Bechstedt F., 
    Phys. Rev. B, 2006, 
   \textbf{73},\\
    045112, \doi{10.1103/PhysRevB.73.045112}.
\bibitem{he2010first}
    He Y., Zeng T., J. Phys. Chem. C, 2010,
    \textbf{114},
    18023--18030, \doi{10.1021/jp101598j}.


\end{thebibliography}
\end{document}